\documentstyle[12pt]{article}
\begin{document}
\vspace{24pt}
\begin{center}
{\large\sc{\bf Electron Structure  Near Abrupt Edges}}
\vspace{14pt}
{\large\sc{\bf in 2DEG under Strong Magnetic Field.}}
\baselineskip=12pt
\vspace{35pt}

I. Barto\v{s}\footnote{permanent address: Institute of Physics,
Academy of Sciences of the Czech Republic, Prague} and B.Rosenstein
\vspace{24pt}

 Institute of Physics \\
Academia Sinica\\
Taipei, 11529\\
Taiwan\\
\vspace{60pt}
\end{center}
\baselineskip=24pt
\begin{center}
{\bf abstract}
\end{center}

Energies and wave functions of edge states in two
dimensional electron gas
are evaluated  for a finite step
potential barrier model. The spectrum, instead of
smooth bending of Landau branches in the vicinity
of the barrier acquires a steplike form;
 unexpected edge plateaus
and significant energy gap reduction take place between the
neighbouring Landau branches above the barrier tops.
The origin of these phenomena is traced. Stability with
respect to modifications is established.
Manifestation of the qualitatively new features of electron
 densities of states
in abrupt confinements through magnetooptical and
nuclear magnetic spin relaxation
effects is proposed.\newline

\vspace{10pt}
PACS numbers:  73.20 At, 73.20 Dx, 73.40 Hm.
\pagebreak

\section{Introduction}

Strong magnetic fields  and progress in heterostructure
quality allowed discovery of
such dramatic phenomena in two dimensional electron gas (2DEG)
like Integer
and Fractional Quantum
Hall Effect \cite{PG}. In the microscopic theory the edge states,
classically represented by electrons skipping in circular segments
along the edges, play
an important role. Many
magnetotransport experiments in two dimensional electron gas have been
qualitatively understood recently by means of a simple edge - state
model \cite{H}.
This model is based on the picture of smooth Landau level bending
by the potential formed by external charges. The intersections
of each Landau level with the Fermi surface create widely
separated narrow edge channels \cite{Ha}.

The semiclassical notion of the Landau level bending, however, is not always
applicable to the electronic structure
in the vicinity of the barrier representing an edge or
an interface (or
random potential). It is definitely not valid for potentials with
large gradients. In
such situations  often used
quantum mechanical model \cite{MDS} is limited to the extreme
 case of an infinite barrier. Recently Chklovskii {\it et al}
\cite{CSG}, using the self-consistent electrostatic approach, showed that
the resulting effective potential should acquire a steplike shape
even if the external potential is smooth.

In this paper, finite step potential is considered to better approximate the
interface or the steplike potential
 just mentioned. The electron structure exhibits
 two rather unexpected phenomena: the
edge plateaus and the Landau gap reduction.
These phenomena are not present in either of previously studied models.

 The interface might be either
 a "boundary" confining 2DEG or an
interface between two different materials. The finiteness of the
 barrier confining 2DEG is especially important, if the confinement is
realized by means of an interface between two similar materials.
Then, in very strong magnetic fields, the magnetic field
induced splitting (between Landau levels)  may be
comparable to the potential barrier height.
For example, in the 2DEG formed at the interface between GaAs and GaAl$_{1-x}$
As$_x$, the effective interface potential barrier is about $0.3\ eV$.
The Landau level spacing in strongest magnetic fields experimentally available
is just a few times smaller. In particular, for the steplike
potential with  wide steps (on the scale of magnetic
length) of height
$\hbar\omega_c$ \cite{CSG,Brey}, the two quantities are equal.

Two dimensional electron systems are realized at abrupt interfaces
 between two slightly different semiconductor crystals. Such systems are
 prepared by molecular beam epitaxy where deposition as well as doping can
be controlled  on atomic layer level.
Lateral confinement within the plane
of 2DEG can be achieved in various ways. In one of them, the gate
voltage restricts the electron motion. Due to the distances between
the gate and the 2DEG plane, the resulting confinement potential
is very smooth within the plane.
Mesa etched samples  in which the surface charge substitutes the gate
voltage, fall also into this category of smoothly confined systems.
These have been studied in detail theoretically \cite{GH}.

Recently, systems with varying chemical composition
along the growth plane have been prepared. In such systems
the effective confinement
barrier height is approximately given by the well studied discontinuity in
valence and conduction bands at heterostructures.  Moreover,
 these barriers are
 localized within
just a few interatomic distances. Therefore on a scale of magnetic length,
even for strong magnetic fields they represent very abrupt barriers.
For such systems our model is appropriate. Systems of this type have
been also prepared in a form of very narrow channels (quantum wires).
In these systems
consequences of interaction between the two opposite edges
can be studied.

The paper is organized as follows.
 In section 2, general properties of confinement barriers are reviewed.
In particular, the concept of lagging of the electron center of mass
behind the Larmor orbital center, due to the edge, is introduced.
Next section is devoted to a rather detailed analysis of the simplest model
of an abrupt barrier: the finite rectangular step. Spectrum exhibits
two rather unexpected phenomena. Landau branches do not rise steadily
towards the edge, but instead develop a series of "edge plateaus".
The energy separation between the Landau branches does not remain constant:
at certain regions within the edge it gets significantly reduced compared to
$\hbar\omega_c$. Both of
these phenomena are peculiar to  finite
abrupt edge barriers and
do not exist in smoothly confined systems.
In section 4, some experiments in which the characteristically modified
electron
densities of states should be observable
like nuclear magnetic spin resonance and magnetooptics are discussed.
The stability of results obtained for the simple model is examined in section
5.

\section{General description of the edge.}

Here, we consider a system of 2D noninteracting electrons in the vicinity of a
boundary under the homogeneous
magnetic field  $B$ perpendicular to the  $xy$ - plane.
The edge or interface is described by a potential barrier $V(
x)$.

The one particle Hamiltonian is \begin{equation}
H=\frac{1}{2m} ({\vec p} - \frac{e}{c} {\vec A})^2+V(x).
\end{equation}
where ${\vec A(x)}$ is the vector potential describing
the applied magnetic field. In the Landau gauge, ${\vec A}\equiv (0,Bx,0)$,
the motion along the y - direction is free and
we can separate variables:
\begin{equation}
\psi_{n,X} (x,y)=\frac{1}{{\sqrt 2\pi}}{\exp \left(\frac{iXy}{l^2}\right)}
\phi_{n,X}(x).
\end{equation}
The wave function $\phi_{n,X}(x)$ obeys the one dimensional Schr\"odinger
equation:
\begin{equation}
\left[-\frac{\hbar^2}{2m}\frac{d^2}{dx^2}+\frac{1}{2}m\omega_c^2(x-X)^2+
V(x)\right]
\phi_{n,X}(x)=E_n(X)\phi_{n,X}(x)
\end{equation}
where $l\equiv  {\sqrt \frac{\hbar c}{eB}}$ is the magnetic length,
 $\omega_c\equiv
\frac {eB}{mc}$ is the cyclotron frequency and $X$ is coordinate of the
center of a Larmor orbit.
This is  the Schr\"odinger equation
for the harmonic oscillator superimposed with the barrier.
The integer
$n$ parametrizes discrete Landau levels, $n=0, 1, 2,...$.

If the barrier $V(x)$ is a smooth function of $x$ (so that the force
$\frac{\partial V(x)}{\partial x}$ is small compared to $\hbar\omega_c/l$),
then all the Landau levels follow the potential:
\begin{equation}
E_n(X)=\left(n+\frac{1}{2}\right)\hbar\omega_c+V(X)
\end{equation}
In this case the spacing between different Landau branches
always remains a multiple of $\hbar\omega_c$.

In another extreme
case of an infinite step barrier,  simple boundary condition
of vanishing of the wave function is  to be imposed:
$\phi_{n,X}(0)=0$.
 It was studied by McDonald and St\v{r}eda \cite{MDS},
who obtained energies of first few levels as a function of
the distance $X$. Deep inside the region to the
left of the barrier ($X<<0$) the influence
of the interface is negligible and $E_(X)\rightarrow \hbar \omega_c \left(n
+\frac {1}{2}\right)$. As $X$ approaches the barrier,
 the energy levels rise
due to repulsive effect of the infinite barrier. For orbits centered
in the "forbidden" region to the right of the barrier,
 the energies continue to rise indefinitely.

Certain general statements about the electron structure can be made without
specifying the shape of the confinement barrier $V(x)$.
The Hamiltonian of the 1D Schr\"odinger equation depends on $X$. The
Hellmann - Feynman theorem \cite{Merzbacher}
 applied to $X$ as a parameter provides
the following useful relation  which is independent of the barrier profile:
\begin{equation}
 \frac {dE_n(X)}{dX}=<\phi_{nX}|\frac {dH_X}{dX}|\phi_{nX}>=m\omega_c^2
<\phi_{nX}|X-x|\phi_{nX}>
\end{equation}
An
 important quantity $X-{\bar x}$, where ${\bar x}$ is the average
electron position, will be referred to as a displacement
of the classical Larmor orbit center due to repulsion by the barrier.
The displacement therefore becomes:
\begin{equation}
 X-{\bar x}=\frac{1}{m\omega_c^2} \frac {dE_n(X)}{dX}\label{dis}
\end{equation}
 The center of mass ${\bar x}$ characterizing the wave function
$\phi_{nX}(x)$ is thus easily obtained by differentiating the dispersion
relation
 $E_n(X)$\footnote{This formula was derived by
O. Heinonen and S. Taylor using variational principle, see \cite{HT}.}.
It is clear, that far from the barrier, $x<<X$, the displacement is zero,
and
$X={\bar x}$. When the Larmor orbital center $X$ approaches the barrier, its
repulsion gives rise to lagging of ${\bar x}$ behind $X$.
 For  the infinite confinement barrier,
 lagging of ${\bar x}$ behind its Larmor orbit's center $X$
 steadily increases as
$X$ grows (the wave function is nonzero for $x<0$ only).

 For any monotonously rising
potential the function $E_n(X)$ is monotonous
in $X$\footnote{This follows from a simple variational argument.}.
  In the particular case of the finite step barrier, ${\bar x}$ laggs behind
$X$ (it will be proved later that for monotonously increasing $V(x)>0$ the
derivative $ \frac {dE_n(X)}{dX}\ge 0$ for any $X$ and $n$).
The magnitude of the displacement, however, for finite barriers,
 may rise and fall
with growing $X$, as we will
see in subsection 3.6..

\section{The elementary rectangular step model.}

Let us now concentrate on the case of abrupt potentials. The basic potential
of this kind is a single rectangular barrier $V(x)=V\theta(x)$. Generalizations
to several steps or to less abrupt barriers will be considered later.
We start with an analytic
 perturbative treatment of small barriers which will be
shown to
contain all the qualitative features of the exact solutions.

\subsection{Approximate calculation.}

If $V<<\hbar\omega_c$ or if a state is not very close to the interface,
we can use the perturbation theory  around Landau levels
to calculate the dispersion
relation $E_n(X)$.
The perturbed energies to the first order in the barrier potential $V$
are:
\begin{equation}
E_n(X)=(n+1/2)\hbar\omega_c + _0<n,X|V\theta(x)|n,X>_0=(n+1/2)\hbar\omega_c +
V\int_0^{\infty}dx\ \rho_n^0(X,x)
\end{equation}
where $\rho_n^0(X,x)$ is  the electron density
corresponding to the solution of eq.(3) without the barrier.
It is given by
$$\rho_n^0(X,x)=\rho_n^{Lan}(x-X)$$
\begin{equation}
\rho_n^{Lan}(x)\equiv \frac{1}{\sqrt{\pi}2^{n}n!\ l}\exp
\left[-\frac{x^2}{l^2}\right]
H_n^2(x/l)
\end{equation}
where $H_n(x)$ are the Hermite polynomials of the order $n$.
The integral can be performed to obtain analytic expressions for the
first order corrections of the energies:
$$\Delta E_0= V \left[1/2+1/2\ {\rm  Erf}(X/l)\right]$$
\begin{equation}
\Delta E_1= V\ \left[1/2+1/2\  {\rm Erf}
(X/l)-\frac {X/l}{2 \sqrt\pi}\exp \left({-X^2/l^2}\right)\right]
\end{equation}
for the first two levels. Differentiating eq.(7) with respect to $X$ one
obtains:
\begin{equation}
\frac{\partial E_n(X)}{\partial X} =V\rho_n^{Lan}(X).
\end{equation}
The inverse density of states of the $n^{th}$ Landau branch is proportional
to $\frac{\partial E_n(X)}{\partial X}$.
Now, since the electron density, eq.(8), has $n$ nodes, perturbatively,
 there are precisely $n$ infinitely flat regions of the dispersion relation.
These will be called edge plateaus.
For example, $E_1(X)$  has such region around $X=0$.
Generally, the zeros of the Hermite polynomials thus determine the
 plateau centers.

 The gap between the ground and the first excited Landau
branches
does not remain $\hbar\omega_c$. For $X>0$ it gets reduced and reaches
its minimum  value
$\hbar\omega_c-\frac{e^{-1/2}}{2^{3/2}{\sqrt \pi}}V$ at $X=l/{\sqrt 2}$.

These
two features turn out to be of general nature rather than
just an artifact
of the perturbation theory. As we show next, they become even more
pronounced for
higher barriers.

\subsection{Spectrum.}

In order to solve  the quantum mechanical problem eq.(3)
with  the
rectangular potential for arbitrary height $V$ beyond perturbation theory,
the  usual quantum mechanical matching of the wave function
has to be performed along the boundary line $x=0$. This was done
for the finite barrier in \cite{Vigneron} (in which, however, the edge plateaus
have not been noted).

The wave function matching for $\phi_{n,X}(x)$
has to be performed at a single point $x=0$.
It is convenient to shift the origin of the coordinate system to $X$. In
 natural
units of magnetic length the new variable $x'$ is:
 $x'\equiv  \frac{\sqrt 2}{l} (x-X)$. With energy
expressed
in units of Landau spacing $\hbar \omega_c$,
$ \nu_n \equiv E_{n,X}/(\hbar \omega_c)-\frac {1}{2}$,
eq.(3) takes a form
\begin{equation}
\left[ \frac {d^2}{dx'^2}-\frac {1}{4} x'^2 - V \theta(x'+{\sqrt 2}X/l) +
\left(\nu_n+\frac {1}{2}\right)
\right]\phi_{n,X}(x')=0
\end{equation}
which is the differential equation defining the parabolic cylinder functions
\cite{Merzbacher,E}. The two linearly independent solutions $D_{\nu_n}(-x')$
 and
$D_{\nu_n-V}(x')$ satisfy asymptotically the conditions of rapid decrease
for $x'\rightarrow -\infty$ and  $x'\rightarrow +\infty$, respectively.

The matching of the logarithmic derivatives at $X'=-{\sqrt 2}X/l$
 (or equivalently
the condition of zero Wronskian in the expression for the Green's
function of the system) gives:
\begin{equation}
\frac {D'_{\nu_n-V}(x')|_{x'=X'}}{D_{\nu_n-V}(X')}-
\frac {D'_{\nu_n}(-x')|_{x'=X'}}{D_{\nu_n}(-X')}=0
\end{equation}
This determines the energy levels $\nu_n$ as functions of the position $X$.
The equation (12) was solved numerically using a simplified form
(obtained from the well known relations expressing derivatives of
 the parabolic cylinder
functions \cite{E}):
\begin{equation}
 D_{\nu_n+1}(X') D_{\nu_n-V}(-X')+D_{\nu_n-V+1}(-X') D_{\nu_n}(X')=0
\end{equation}

The dispersion relations for electrons $E_n(X)$ (Landau branches)
in the vicinity of the barrier have been
evaluated numerically. Results for $V=\hbar\omega_c$ and $V=5\hbar\omega_c$
are given in Fig.1 and Fig.2\footnote{Results for a high barrier
  $V=5\ \hbar\omega_c$
were reported earlier \cite{BR}.}. The ground Landau branch behaves as
expected. It starts at its left asymptotic value $\nu=0$ and gradually
rises to the right asymptotic value $V+\nu$.
The  spectrum  of excited Landau branches above the
top  of the barrier is rather surprising, however.
 Instead of smooth transition from
one asymptotic region ($X<<0$) to another ($X>>0$), a steplike rise
 is obtained. Two unexpected features can be clearly seen:\newline
(i) {\it  Edge plateaus.}\newline
The $n^{th}$ Landau branch $E_n(X)$ has $n$ very pronounced  edge
plateaus.
We call these edge plateaus  in order
to differentiate them
from  those in the asymptotic regions
(the bulk plateaus). Density of states at the edge plateau energies
is strongly enhanced.\newline
(ii) {\it  Landau gap reductions.}\newline
The $n^{th}$  Landau branch
almost touches  the $(n-1)^{th}$  one
 $n$ times.  Unlike in
the case of smooth confinement in which all the Landau branches
follow the underlying potential \footnote{Similar behaviour is observed in
the opposite extreme case of abrupt and infinite barrier \cite{MDS}.},
the energy gap between two neighbouring branches exhibits regions
of significant reduction, in particular for higher barriers.

Following a Landau branch edge plateaus and the gap reduction regions
alternate. First we would like to understand the origin of
the edge plateau phenomenon.

\subsection{Why there are infinitely flat plateaus for the step potential?}

The plateaus in Fig 1 and 2 look, in fact, very flat. How flat? In order to
see this, we calculate the density of states $D(E)$. As it is well known
\cite{H} the density of states in 1D system is just an inverse of
 the derivative
\begin{equation}
 D_n(E)=\frac {1}{2\pi}\left[\frac {dE_n(k_y)}{dk_y}\right]^{-1}=
\frac {1}{2\pi l^2}\left[\frac {dE_n(X)}{dX}\right]^{-1}
\label{dos}
\end{equation}
Fig.3 shows the density of states of the third excited $n=3$ Landau
level for $V=5\hbar\omega_c$.
The three peaks around $\nu=5.7$, $\nu=6.4$ and $\nu=7.1$ indicate
infinite density of states. In Fig.4, full
density of states from all the branches below Fermi level
$E_F=3.5\ \hbar\omega_c$ is plotted.

Now  we will show that this is indeed the case for any rectangular potential.
We apply the Hellmann - Feynman theorem  to Hamiltonian
eq.(11), again with respect to $X$, to obtain:
\begin{equation}
 \frac {dE_n(X)}{dX}=
<\phi_{nX}|V\delta(x'+X)|\phi_{nX}>=V|\phi_{nX}(x'=-X)|^2
\end{equation}
Returning back to the variable $x$ the result is
\begin{equation}
\frac {dE_n(X)}{dX}=V\phi_{nX}^2(x=0)
\label{d}
\end{equation}
This is the square of the (real) wave function at the
interface. Note that from eq.(\ref{d}) we can obtain the
 monotonicity of $E_n(X)$\footnote{It is convenient to use eq.(\ref{d})
 to normalize
the wave functions. This circumvents the numerical integration.}

THI FILE ARRIVED TRUNCATED. THE AUTHORS DIDN'T CARE. NEITHER DO WE.

\end{document}